\documentclass[journal,twoside,web]{ieeecolor}
\usepackage{generic}
\usepackage{cite}
\usepackage{amsmath,amssymb,amsfonts}
\usepackage{algorithmic}
\usepackage{graphicx}
\usepackage{textcomp}
\usepackage{hyperref}
\usepackage{bm}
\usepackage{amsmath}
\usepackage{booktabs}

\def\BibTeX{{\rm B\kern-.05em{\sc i\kern-.025em b}\kern-.08em
    T\kern-.1667em\lower.7ex\hbox{E}\kern-.125emX}}
\begin{document}

\title{Advancing 3D Medical Image Analysis with Variable Dimension Transform based Supervised 3D Pre-training}
\author{Shu Zhang, Zihao Li, Hong-Yu Zhou, Jiechao Ma and Yizhou Yu, \IEEEmembership{Fellow, IEEE}\\
\thanks{This work is funded by the National Key Research and Development Program of China (No. 2019YFC0118101) and Zhejiang Province Key Research \& Development Program (No. 2020C03073).}
\thanks{Shu Zhang and Zihao Li contributed equally. Yizhou Yu is the Corresponding author.}
\thanks{Shu Zhang, Zihao Li and Jiechao Ma are with the Deepwise Artificial Intelligence Laboratory, No.8, Haidian Avenue, Haidian District, Beijing, P.R.China 100080 (e-mail: zhangshu@deepwise.com;lizihao@deepwise.com;majiechao@deepwise.com).}
\thanks{Hong-Yu Zhou and Yizhou Yu are with the Department of Computer Science, The University of Hong Kong, Pokfulam, Hong Kong (email: whuzhouhongyu@gmail.com;yizhouy@acm.org).}}

\maketitle

\begin{abstract}
The difficulties in both data acquisition and annotation substantially restrict the sample sizes of training datasets for 3D medical imaging applications. As a result, constructing high-performance 3D convolutional neural networks from scratch remains a difficult task in the absence of a sufficient pre-training parameter. Previous efforts on 3D pre-training have frequently relied on self-supervised approaches, which use either predictive or contrastive learning on unlabeled data to build invariant 3D representations. However, because of the unavailability of large-scale supervision information, obtaining semantically invariant and discriminative representations from these learning frameworks remains problematic. In this paper, we revisit an innovative yet simple fully-supervised 3D network pre-training framework to take advantage of semantic supervisions from large-scale 2D natural image datasets. With a redesigned 3D network architecture, reformulated natural images are used to address the problem of data scarcity and develop powerful 3D representations. Comprehensive experiments on four benchmark datasets demonstrate that the proposed pre-trained models can effectively accelerate convergence while also improving accuracy for a variety of 3D medical imaging tasks such as classification, segmentation and detection. In addition, as compared to training from scratch, it can save up to 60\% of annotation efforts. On the NIH DeepLesion dataset, it likewise achieves state-of-the-art detection performance, outperforming earlier self-supervised and fully-supervised pre-training approaches, as well as methods that do training from scratch. To facilitate further development of 3D medical models, our code and pre-trained model weights are publicly available at \href{https://github.com/urmagicsmine/CSPR}{https://github.com/urmagicsmine/CSPR}.

\end{abstract}

\begin{IEEEkeywords}
3D Medical Image, Transfer Learning, Variable Dimension Transform, Supervised Pre-training
\end{IEEEkeywords}


\section{Introduction}
\label{sec1}

With the rapid advancement of deep learning techniques, fast and accurate medical image analysis systems have emerged as an instrumental tool for routine clinical practice. These systems have been deployed with the objectives of improving efficiency and accuracy in a variety of applications, including but not limited to assisting radiologists in image interpretation with automatic lesion detection~\cite{liu2020cross,ren2016faster}, improving the accuracy of prognostic evaluation or disease triage with image classification~\cite{peng2019prognostic,wu2020covid,he2016deep}, and improving the efficiency and accuracy of target area delineation in radiotherapy with automatic target area segmentation~\cite{kamnitsas2017efficient,falk2019u}. Deep learning algorithms show promising results in the medical field, similar to their success on natural images; nevertheless, large-scale annotated medical image datasets are still required to develop deep learning models further.

Collecting sufficient annotated medical image data remains a considerable difficulty when compared to building large-scale annotated natural image datasets with millions or even billions of annotations.  On the one hand, medical data collecting is strictly regulated for privacy reasons; on the other hand, the requisite specialist knowledge and the tedious nature of medical image annotation make the formation of new annotated medical datasets prohibitively expensive and time-consuming. As a result, transfer learning employing weights pre-trained on large-scale 2D natural image datasets (e.g. ImageNet ~\cite{ILSVRC15}) has become the de-facto paradigm for speeding up model convergence and improving overall model performance on small-scale medical image datasets. Despite the fact that there is a domain shift between natural images and medical images, in reality, representations learned from natural images can significantly improve feature discrimination ability for 2D medical image analysis tasks ~\cite{gulshan2016development,lakhani2017deep}. 

At now, pre-training approaches are largely developed for 2D CNNs. Not until recently, more researches have attempted to solve this issue for 3D CNNs, i.e., proposing self-supervised or fully-supervised models ~\cite{zhou2020models,zhu2020rubik,gibson2018niftynet,chen2019med3d} to pre-train a universal 3D model, which can be subsequently utilized to fine-tune 3D CNNs on the target task. Furthermore, because non-annotated 3D medical data is on a much bigger scale than annotated data, a rising amount of research attention has been paid to learning self-supervised 3D representations by building various proxy tasks on top of unannotated medical datasets. These proxy tasks, which include but are not limited to anatomical similarity models ~\cite{zhou2020models} and cube ordering~\cite{zhu2020rubik}, are frequently reliant on the use of prior knowledge information. They deeply mine free supervision signals from unlabeled data, and empirical results show that they outperform most 2D/2.5D approaches as well as 3D models trained from scratch~\cite{li2019mvp,ni2019elastic}.

\begin{table*}[!tbp]
    \centering
    \setlength{\belowcaptionskip}{5pt}
    \caption{Different 3D representation learning methods and their characteristics.}
    \label{Datasets}
    \begin{tabular}{cccccc} 
    \hline
        Pre-train Type & Method & Source modality & Dataset Scale & Cross Domain & Training Task  \\ \hline
        Self-supervised & Rubik's Cube & CT &  Medium & N & Restoration\\ 
        Self-supervised & Models Genesis  & CT & Medium & N & Restoration \\ \hline
        Supervised & NiftyNet & CT &   Small & N & Segmentation \\ 
        Supervised & Med3D & CT,MRI &   Medium & N & Segmentation \\ \hline
        Supervised & Kinetics & Video Data &  Large & Y & Classification \\  
        Supervised & I3D      & Natural Image &  Very Large & Y &  Classification \\  \hline
    \end{tabular}
\end{table*}

Nonetheless, despite their effectiveness, unsupervised learning algorithms, including self-supervised learning, are frequently overshadowed by supervised learning algorithms. For example, as demonstrated in the most recent 3D pre-training studies \cite{zhou2020models,zhou2020comparing}, even state-of-the-art self-supervised learning algorithms could not outperform ImageNet pre-trained models. This might be because, due to a lack of supervised signals, semantically discriminative representations are difficult to be mined from un/self-supervised learning algorithms. Unfortunately, there are no publicly accessible annotated 3D medical image datasets that are large  and diverse enough to give adequate training data and supervision signals for universal 3D feature learning. \cite{chen2019med3d} proposed to build a large 3D heterogeneous medical dataset with diverse modalities, target organs, and pathologies, and designing a heterogeneous Med3D network to co-train on the multi-domain dataset. However, the generated dataset is still too small to be useful for pre-training 3D CNNs.

In this paper, we introduce a simple yet effective \textit{\textbf{S}upervised pre-training technique based on \textbf{V}ariable \textbf{D}imension transform} (\textbf{SVD}-Net). The proposed SVD-Net is employed to learn 3D representations from large-scale 2D natural image datasets, which are then used for transfer learning in target 3D medical tasks. The data-scarce problem is addressed with our proposed variable dimension transform, which reformulate 2D natural images to yield our source of 3D data. During the process, the color information is transformed to pseudo-3D structure information, which is then used to learn 3D structural and textural representations. In contrast to self-supervised approaches, we might employ high-level vision tasks as proxy tasks to benefit from semantic supervision for learning discriminative and invariant 3D representations.

The idea of pre-training a 3D network with 2D natural images was initially proposed as a trick in our conference paper~\cite{zhang2020revisiting}, the objective of which is to construct 3D models for universal lesion detection. This work significantly expands on the preceding conference paper in the following aspects: Firstly, we revisit the mixed usage of RGB images and adjacent slices of 3D volume images in medical problems to shed lights on the root of our proposed idea, and further re-establish the idea as the variable dimension transform based supervised pre-training technique. Secondly, this work highlights a comprehensive experimental analysis to confirm the efficacy of the proposed 3D pre-training approach. To validate its universal effectiveness on 3D medical image analysis tasks, we conduct experiments on four benchmark 3D medical datasets and compare our proposed pre-trained 3D models to existing state-of-the-art self-supervised and fully-supervised pre-trained models on four target tasks, namely, pulmonary nodule classification, pulmonary nodule segmentation, liver segmentation, and lesion detection. The SVD-Net, as simple as it appears, is quite effective in practical applications, supporting the principle of translating color information into 3D structural and textual information for efficient feature learning. Last but not least, comprehensive experiments are designed to investigate the design decisions that contribute to the success of the compact SVD-Net.

\section{Related work}


In clinical practice, the inspection of 3D medical data requires information on 3D shape and texture. Nevertheless, popular 2D deep learning based methods, such as\cite{roth2014new}, can only capture spatial correlation while leaving the rich 3D context information unexploited. The use of 3D convolution is the most straightforward solution for effective 3D context modeling. 
However, compared to 2D CNNs, training deep 3D CNNs from scratch is much more difficult and prone to overfitting due to the increased number of parameters. Therefore, the research of 3D pre-trained models are of significant importance for its development.

\begin{figure*}[!tbp]
    \centering
	\includegraphics[width=6in]{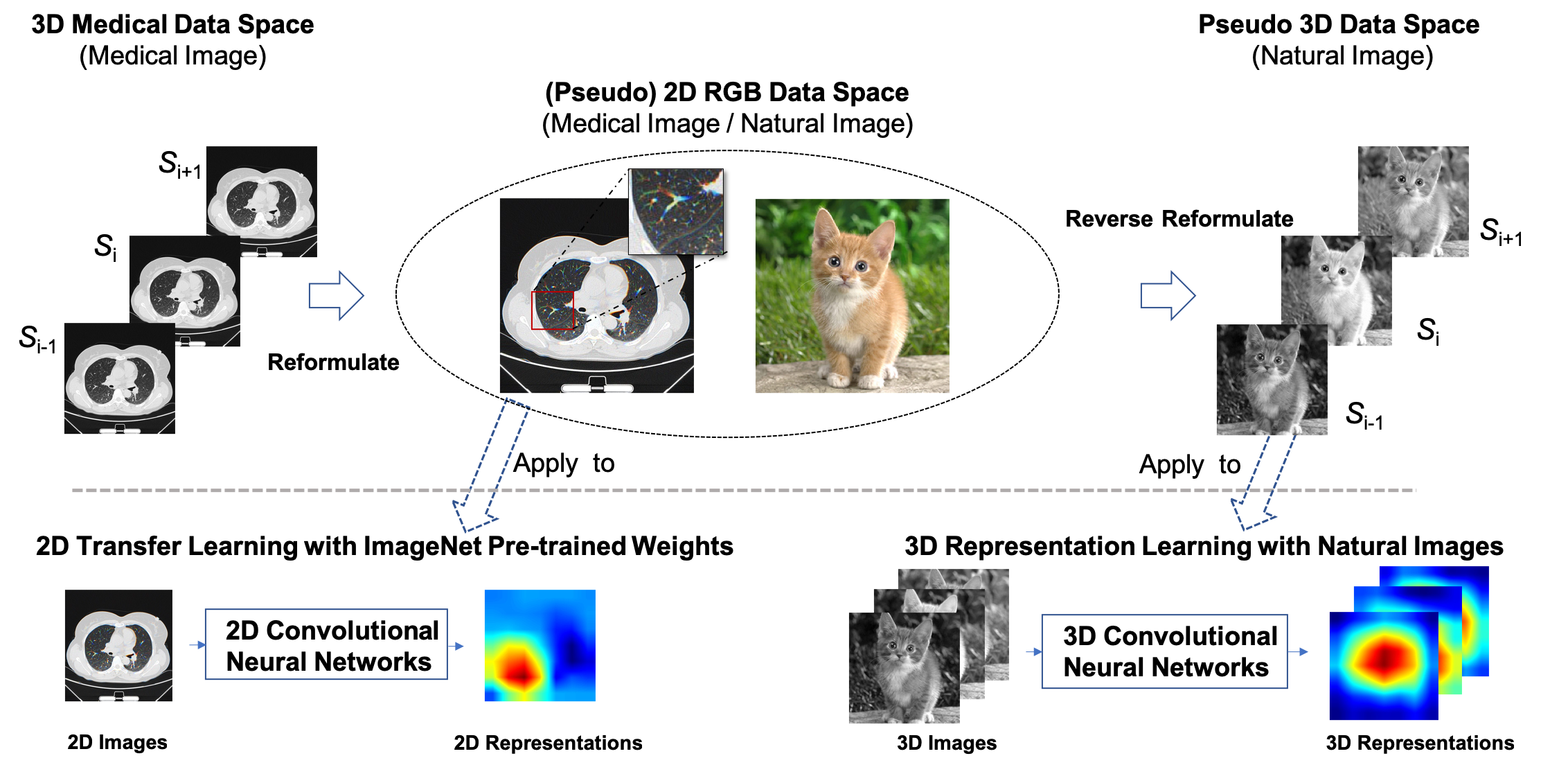}
	\caption{Illustration of the Variable Dimension Transform. 
	To take advantage of model weights pre-trained on ImageNet, 3D medical images are often reassembled to simulate 2D natural images. On the contrary, we reversely reformulate 2D natural images to the form of 3D medical images for 3D supervised pre-training. During the variable Dimension transform process, color information in the original 2D images are transformed to pseudo 3D structure information and would be exploited by 3D CNNs for effective feature learning. Best viewed in color and with zoom.}
	\label{reform}
\end{figure*}

\subsection{Transfer Learning from 2D Weights}

To alleviate the problem of the lack of 3D pre-trained models, some studies explored to generate 3D weights from 2D ImageNet pre-trained models for further transfer learning. I3D\cite{carreira2017quo} proposed to repeat 2D convolution kernels along the new axis for $k$ times to form the $K\times K\times K$ 3D convolution layers. Similarly, \cite{shan20183} extended the 2D pre-trained convolution kernels to 3D with the additional parameters initialized by the means of zero padding.
However, these methods failed to model 3D context in the z-axis with the repeated 2D kernels. It can only be adopted to capture spatial information from multiple axial slices.
To make up for the inconsistent representation ability along the depth axis and the height/width axis, ACS\cite{yang2019reinventing} further performed a symmetric pseudo-3D convolution operation. It expanded the dimensions of 2D convolution kernel along depth, height and width axis respectively to generate three view-based 3D convolution kernels. Input feature maps are divided into three parts and view-based 3D convolutions are performed on each part. 
With such a design, 2D weights can be seamlessly copied to the newly generate view-based 3D kernels and the resultant 3D convolutional layers have the ability to model structure and texture information in all three views. However, the kernel-by-kernel conversion might break the correlation among subsequent layers and thus potentially hurt the performance. Therefore, there are more methods that explore to learning 3D representations as a whole network rather than converting existing pre-trained 2D weights kernel by kernel separately.


\subsection{Fully-supervised 3D Representation Learning Methods}

NiftyNet~\cite{gibson2018niftynet} provided a modular pipeline for a range of medical imaging applications as well as a manually annotated abdominal CT dataset, which contains 90 abdominal CTs collected from publicly available datasets. 
There are in total 8 kinds of organs annotated by radiologists with pixel-level annotation. The scale of this dataset is relatively small and is not sufficient for effective training of large 3D convolution models. Considering the difficulty of data collection and manual annotation, it's impractical to acquire a single large-scale, high-quality 3D image dataset for 3D CNN pre-training. To address this issue, \cite{chen2019med3d} aggregated eight datasets from several medical image challenges to build a larger heterogeneous dataset. They further proposed a heterogeneous 3D network called Med3D to obtain a series of pre-trained CNN models with supervised tasks like segmentation. 
The Med3D enjoys the advantage of a similar data distribution with a number of target tasks, but its small data scale severely limits the performance of the pre-trained models.

\subsection{Self-supervised 3D Representation Learning Methods}

One of the common strategies for self-supervised learning is generating artificial data and label pairs by constructing an image proxy task (e.g. image restoration), and training the network in a fully-supervised manner to learn good representations. 

Authors in \cite{chen2019self} proposed to learn semantic features by a context restoration proxy task which randomly picking some boxes in medical images and exchanging their content. \cite{zhou2020models} further explored more image transformations including non-linear pixel value mapping, local pixel shuffling, image inner and outer cutouts, and built a set of pre-trained models nick-named Models Genesis for different medical imaging tasks. \cite{haghighi2020learning} extended Models Genesis to conduct Semantic Genesis which leveraged semantically enriched visual representations from the consistent and recurrent anatomical patterns. 

Authors in \cite{zhuang2019self} presented Rubik's Cube, their proxy task is to predict the order and rotation state of the 3D sub-cubes cropped from the entire 3D data cube. \cite{zhu2020rubik} added an extra cutout restoration task based on Rubik's Cube to achieve better generalization ability. 
Furthermore, \cite{tao2020revisiting} adopted a GAN-like structure to recover the original state of Rubik's cube from the disarranged state so as to better exploit the inherent 3D anatomical information of organs. 

In Table\ref{Datasets}, we give a brief summary of some of the aforementioned pre-trained models according to their data modalities, dataset scales and types of proxy task.

\section{Methods}

To address the deficiency of training 3D CNNs from scratch, in this paper, we propose to pre-train a high-performance 3D CNN by leveraging large-scale supervised learning tasks in the natural image domain. 
The obtained generic and powerful 3D representations can be further used as pre-trained weights to boost the performance of target 3D medical image analysis tasks through transfer learning (model fine-tuning). 

Details of how our SVD-Net are trained and transferred will be elaborated in the following subsections. Two key questions regarding fully-supervised 3D pre-training using natural images are addressed, i.e. how to construct large-scale 3D data from natural images and how to design 3D network architectures used in such pre-training.


\subsection{Variable Dimension Transform}
\label{data_reform}

Deep neural network models pre-trained in the natural image domain have been successfully explored for medical imaging tasks. For instance, for 3D medical imaging problems, researchers often combine three adjacent slices to form a three-channel RGB image~\cite{yan20183d, li2019mvp} so that weights pre-trained on ImageNet can be exploited for transfer learning.
The left part of Figure~\ref{reform} illustrates the process of converting adjacent slices in CT scan to RGB images. For most well-aligned structures, their appearance after reassembling resembles the original slice input. As for pulmonary vessels, the change in their position and shape among neighboring slices in the original 3D space is transformed into color information in the RGB space. Thus, after such a reformulation, rich 3D structural information is preserved in the 2D space through color encoding. For such reformulated 2D data, color encodes completely different information from that of the original natural images. Nonetheless, it can still benefit from ImageNet pre-trained weights to achieve better performance.

Intuitively, the 3D-to-2D reformulation process motivates us to conduct a reverse reformulation, where we call it the variable dimension transform. It decomposes a natural image with RGB channels into three adjacent slices in the 3D space. As shown in the right part of Figure~\ref{reform}, given a 2D natural image $X \in (C_{in}, H_{in}, W_{in})$ where $C_{in}=3$ denotes the three channels of the RGB image, we reshape it as $( C', D', H{in}, W{in})$ where $D'=C_{in}=3$ and $C'=1$ to form a pseudo 3D input. 
Through such reverse reformulation, we convert the color information in the original 2D natural image into complex 3D structural information in the simulated 3D image. 3D convolution kernels learned on such pseudo-3D data can potentially model complex 3D structures and textures that exist in 3D medical images.

With the proposed variable dimension transform, we can construct large-scale 3D datasets for fully-supervised 3D pre-training as the natural image domain has by far the largest annotation scale for visual learning. 
Supervision signals in the large-scale annotated datasets, which are the fundamental driving force for learning generic and powerful 3D representations, can be exploited for 3D feature learning.

\begin{figure}
    \centering
 	\includegraphics[width=3in]{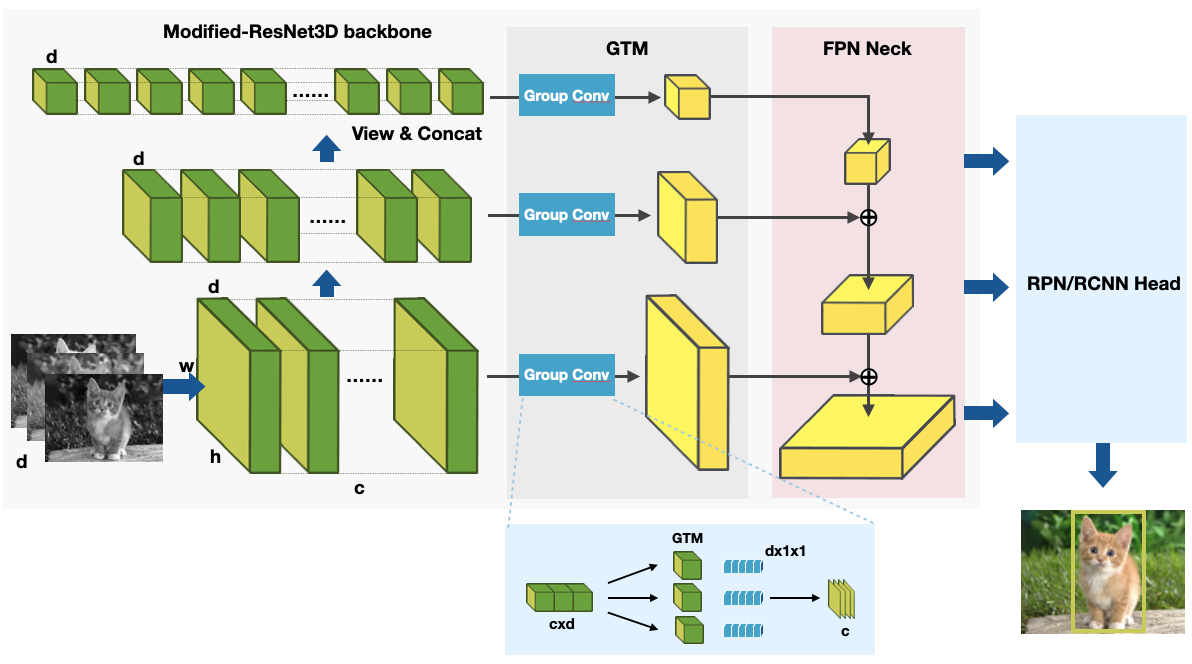}
	\caption{Network architecture for pre-training detection models on the MS-COCO Dataset. The proposed architecture has a 3D backbone and a 2D FPN neck, a 2D RPN and an RCNN head. The input image is reformulated as 3D data, while the object-level bounding box annotations remain 2D. The GTM is adopted to transform 3D features to 2D for further prediction. }
	\label{arch}
\end{figure}


\subsection{Architecture Design and Supervision Tasks}

We leverage existing popular backbone architectures for feature extraction and focus on training such backbone networks for 3D transfer learning. When integrated with proper necks (e.g. FPN~\cite{lin2017feature}), decoders (e.g. U-Net~\cite{falk2019u}, PSPNet~\cite{zhao2017pyramid}) or heads (R-CNN head~\cite{ren2016faster}), these backbone networks can be further extended to carry out detection, segmentation and classification tasks.

Off-the-shelf 3D backbones perform down-sampling along all $D, H, W$ axes after layers of convolutions, while for our pseudo-3D input, the size of the depth dimension is far less than that of the height and width dimensions. Simply applying the original down-sampling operation would degenerate the depth axis to one single slice, which would apparently neutralize the learning ability of 3D context modeling. 
To tackle this problem, we keep the resolution of the depth dimension fixed by setting all the down-sampling ratios (pooling or convolution stride) along the depth axis to $1$, while retaining original ratios along the height and width axes. In the meantime, zero padding with the size of $(kernel\_size - 1)/2$ is adopted along the z-axis to preserve the depth of the 3D feature maps.
The shape of the output from the modified 3D CNN is $(D_{out}, C_{out}, H_{out}, W_{out})$ with $D_{out}$ kept as 3.

\noindent{\textbf{Pre-training with Classification Task:}}
Benefiting from such a modified architecture, we can make use of a large number of annotated natural images for fully supervised pre-training. For example, when the pre-training task is the ImageNet classification task, after obtaining the 3D feature map, we can use a 3D global average pooling layer to aggregate the feature vector, and employ a fully-connected layer afterwards to produce the final category prediction for the input image. Denote the input image and its category label as $x$ and $y$, and the modified 3D classification network as $\theta$, we adopt cross entropy as the loss function, which can be described as follows:
\begin{equation}
\label{equ:ce_loss}
\bm{L_c} = -\sum_iy\log(\theta(\bm{x})).
\end{equation}

\noindent{\textbf{Pre-training with Detection Task:}}
The ImageNet dataset was collected from the Internet, and each image in the dataset has been associated with a human-annotated category label. Due to the diversity of data sources, a significant number of images have more than one object categories, while only one of them was taken as the image category label and the rest was ignored. This would force the models to learn partial image features only, and may hurt performance on dense prediction tasks, e.g. object detection. There are similar conclusions~\cite{majurski2019cell} suggesting that taking model weights pre-trained on appropriate tasks as initial weights provides better performance.

In this paper, we also consider object detection on another widely used dataset, MS-COCO, which contains millions of object-level annotations, as a supervised pre-training task to explore the influence brought by different proxy and target tasks. Figure~\ref{arch} illustrates the modified FPN architecture used for pre-training with a detection task. As in a classification task, the backbone network also extracts 3D feature maps from the reformulated natural images.

To accommodate for predicting of 2D ground-truth bounding box annotations, we need to convert the 3D feature maps back to 2D ones before feeding them to the RPN and RCNN heads for further prediction. Specifically, we introduce a group transform module (GTM) to aggregate 3D feature maps into 2D ones. For instance, for a 3D feature map with size $(C_{out}, 3, H_{out}, W_{out})$, we first view the 3D feature map as a 2D one with size $(3C_{out}, H_{out}, W_{out})$, and then apply a group convolutional layer with $C_{out}$ groups (each group contains 3 channels) to fuse features from all neighboring slices to yield the final 2D feature maps $(C_{out}, H_{out}, W_{out})$. 
Afterwards, a 2D FPN and a detection head are appended to its output to produce the final 2D predictions.

Generalizing our proposed method to pre-training with a segmentation task (e.g. Pascal VOC~\cite{everingham2015pascal} and Cityscapes~\cite{cordts2016cityscapes}) is straightforward. However, due to the limited data scale of existing segmentation datasets, we do not adopt any segmentation task as a supervised pre-training task in our experiments.

\subsection{Transferring to Medical Imaging Tasks}
\label{trans}

Once we have obtained the pre-trained network with modified architecture, it is convenient to transfer it to various 3D medical imaging modalities and tasks. We could simply initialize a vanilla ResNet3D network with the pre-trained weights from the modified backbone by copying corresponding parameters form the Conv or BatchNorm layers and neglecting pooling or stride configurations in the modified backbone. 3D down-sampling operations should be configured to be suitable for target tasks. Then according to the type of the target task, we simply add a randomly initialized neck, decoder or head to the backbone. 

It is worth noting that our pre-training framework is both model-independent and task-independent. One can pre-train an arbitrary 3D CNN architecture on any 2D annotated image dataset, which makes the proposed method more flexible and generalizable.

\section{Experiments}


In this section, we first report the performance of the proposed pre-training method on proxy tasks of 2D natural image datasets.
In the next part, we conduct extensive experiments on different target tasks including classification, segmentation and detection, to analyze the effectiveness and generalization of our 3D pre-training method.

\subsection{Pre-training on Natural Image Dataset}
\subsubsection{Dataset and Implementation Details}
\noindent{\textbf{Dataset:}}
The ImageNet~\cite{ILSVRC15} classification dataset provided by ImageNet Large Scale Visual Recognition Challenge (ILSVRC) contains around 1.2 million natural images which belong to 1000 classes.
The images were collected from the Internet and labeled by humans manually. MS-COCO~\cite{lin2014microsoft} is the most widely used dataset for general object detection. It is composed of 118k natural images and a total of 0.9 million bounding boxes annotations for 80 categories. 

Both datasets provide large varieties in data distribution and include rich semantic supervision for learning discriminative and invariant representations. 

\noindent{\textbf{Implementation Details:}}
For 3D network pre-training, we adopt ResNet with 3D convolution with our proposed modifications as the backbone. 2D images are transformed to pseudo-3D format using variable dimension transform as its input. For pre-training with the ImageNet classification task, we first resize images to $3\times1\times256\times256$ and then randomly crop them to the size of $3\times1\times224\times224$. Models are trained with the cross entropy loss for 100 epochs. 
The learning rate is initially set to 1$e$-1 and decayed with the cosine annealing policy. Besides, we adopt label smoothing~\cite{szegedy2016rethinking} for better generalization ability.

For pre-training with MS-COCO, the proposed modified FPN architecture is adopted. We perform horizontal flipping as well as multi-scale training with scales of $(384, 448, 512, 576, 640)$ as data augmentation. Since the CNN parameters are randomly initialized, we apply longer training steps as described in~\cite{he2019rethinking} to ensure model convergence. The models are trained for 72 epochs with an initial learning rate of 2$e$-2, which is further decreased by a factor of 10 after 48 and 66 epochs. The batch size on each GPU is set to 2 due to limited GPU memory, and we use group normalization to replace batch normalization to stabilize training. Besides the ResNet3D architecture, we also pre-train a P3D~\cite{qiu2017learning} based 3D FPN model to acquire computational and memory-efficient 3D context modeling.

Both pre-training tasks are implemented using the \textit{open-mmlab toolbox}\footnote{https://github.com/open-mmlab}. Stochastic Gradient Descent (SGD) with a momentum of 0.9 and weight decay of 0.0001 is adopted as the optimizer.


\subsubsection{Performance Evaluation on the Proxy Tasks}

With the proposed data reformulation and architecture modification, we are able to train 3D ResNets on 2D natural images.
Performances of our proposed 3D networks pre-trained on ImageNet, along with detailed time and space complexity information are presented in Table~\ref{Pretrain}. 
Classification performance of baseline 2D networks are also reported for a fair comparison. 
It's worth noting that all the 3D models get higher top-1 and top-5 accuracy compared with their corresponding 2D counterparts under the same training settings. 
This suggests that the proposed variable dimension transform doesn't harm the performances on proxy tasks; on the contrary, benefiting from enhanced representation power from an additional dimension, the proposed 3D models can yield better performance given the same network depth.
For pre-training on MS-COCO, our detection model with ResNet3D-18 achieves a mean average precision (MAP) of 34.2, while our detection model with ResNet3D-50 and P3D-63 achieves comparable detection performance (37.0 vs 36.5). 


\begin{table}
    \centering
    \footnotesize 
    \renewcommand{\arraystretch}{1.2}
    \setlength{\belowcaptionskip}{5pt}
    \caption{Number of Parameters, Flops and ImageNet pre-training performance of different model architectures.}    
    \label{Pretrain}
    \begin{tabular}{lcccc}
        \toprule
        Models &  Params(M) & Flops(G) & Top-1 (\%) & Top-5 (\%) \\
        \midrule
        ResNet-18  & 11.69 & 1.82 & 70.07 & 89.44 \\ 
        ResNet-34  & 21.80 & 3.68 & 73.85 & 91.53 \\ 
        ResNet-50  & 25.56 & 4.12 & 76.55 & 93.15 \\
        \midrule
        Our ResNet3D-18 & 33.67 & 15.99 & 73.43 & 91.31 \\ 
        Our ResNet3D-34 & 63.98 & 32.65 & 76.11 & 92.71 \\ 
        Our ResNet3D-50 & 48.20 & 23.93 & 77.45 & 93.67  \\ 
        \bottomrule
    \end{tabular}
\end{table}

\begin{table}
    \centering
    \setlength{\belowcaptionskip}{5pt}
    \caption{Dice coefficient on LIDC-IDRI segmentation task. Results are given as $mean\pm standard \ deviation$ format.}
    \label{LIDC-SEG}
    \footnotesize
    \setlength{\tabcolsep}{4mm}{
    \begin{tabular}{lccc}
    \toprule
        Methods & Backbone & Dice & P-value\\ \midrule
        Scratch & ResNet3D-18 & $75.12\pm0.23 $ & 0.00\\
        Med3D & ResNet3D-18 &  $ 75.29 \pm 0.16 $ & 0.00 \\
        Models Genesis & U-Net 3D &  $74.67 \pm 0.28$ & 0.00 \\
        ACS scratch & ResNet3D-18 & $ 74.55+0.35 $  & 0.00 \\
        ACS & ResNet3D-18 & $75.65 \pm 0.22$ & 0.06 \\
        I3D & ResNet3D-18 & $ 74.98 \pm 0.38$ & 0.00 \\
        Kinetics & ResNet3D-18 & $75.62 \pm 0.17 $ & 0.02\\ \midrule
        Ours & ResNet3D-18 & \textbf{$\textbf{75.95} \pm \textbf{0.16}$} & $-$ \\
        \bottomrule
    \end{tabular}
    }
\end{table}

\subsection{Transfer to 3D Medical Imaging Tasks}

By pre-training on either ImageNet or MS-COCO, we are able to obtain 3D backbone networks for transfer learning. In the next few sections, to demonstrate the generalization ability and effectiveness of the proposed pre-trained models, we will evaluate our pre-trained models on three fundamental medical scenarios: 3D medical data classification, segmentation and detection with four specific medical imaging problems in the following three sub-sections.

As have been explained in Sec~\ref{trans}, the pre-trained 3D backbones can be conveniently adapted to new imaging tasks.
To train the newly generated models on target tasks, we can either fine-tune all the layers in the network or keep some of the lower-level layers fixed and only fine-tune the subsequent layers. In all our implementations, we fine-tune all the layers in the network unless otherwise specified.

To provide a thorough model comparison, we include several state-of-the-art 3D pre-training methods in our experiment as well as the baseline of training 3D networks from \textbf{Scratch}. For self-supervised methods, we mainly compare our models with \textbf{Models Genesis}\footnote{Models Genesis: \href{https://github.com/MrGiovanni/ModelsGenesis}{https://github.com/MrGiovanni/ModelsGenesis}}~\cite{zhou2020models}, the most recent state-of-the-art for self-supervised representation learning. For fully-supervised methods, \textbf{Med3D}\footnote{Med3D: \href{https://github.com/Tencent/MedicalNet}{https://github.com/Tencent/MedicalNet}}~\cite{chen2019med3d} and 3D ResNet pre-trained on \textbf{Kinetics}\footnote{Kinetics: \href{https://github.com/kenshohara/3D-ResNets-PyTorch}{https://github.com/kenshohara/3D-ResNets-PyTorch}}~\cite{carreira2017quo} are included for comparison, both of which show impressive performance and have released their pre-trained weights to the public. For methods that convert 2D pre-trained weights, we compare our models with the classic \textbf{I3D}~\cite{carreira2017quo} and the newly proposed \textbf{ACS}\footnote{ACS: \href{https://github.com/M3DV/ACSConv}{https://github.com/M3DV/ACSConv}}~\cite{yang2019reinventing}, which have achieved competitive performance on many 3D medical imaging tasks~\cite{cai2020deep}. For all the compared methods, we use their publicly released model architecture and pre-trained weights in all our experiments. All compared models adopt the ResNet based architecture, except for Models Genesis, whose official pre-trained models are based on a 3D U-Net architecture. If not otherwise specified, ResNet models pre-trained on ImageNet are adopted for evaluation of our proposed method in all the following experiments.

\begin{table*}
    \centering
    \setlength{\belowcaptionskip}{5pt}
    \caption{AUC on LIDC-IDRI classification task. Results are given as $mean \pm standard \  deviation$ format.}
    \label{LIDC-CLS}
    \footnotesize
    \setlength{\tabcolsep}{4mm}{
    \begin{tabular}{lccccccc}
    \toprule
        Methods                       & Accuracy   & AUC        & F1-score         & Precision  & Recall/Sensitivity     & Specificity & P-value\\ 
        \midrule
        Scratch 	& 85.10 $\pm$ 1.56  & 92.75 $\pm$ 0.58  & 71.01 $\pm$ 1.65  & 59.68 $\pm$ 3.52  & 88.13 $\pm$ 3.64  & 84.31 $\pm$ 2.74  & 0.00 \\
        Med3D	& 83.81 $\pm$ 2.68  & 86.50 $\pm$ 1.06  & 65.68 $\pm$ 2.73  & 59.59 $\pm$ 6.65  & 74.37 $\pm$ 5.28  & 86.26 $\pm$ 4.57  & 0.00 \\
        Models Genesis& 86.32 $\pm$ 1.23  & 94.18 $\pm$ 0.43  & 73.14 $\pm$ 1.65  & 61.71 $\pm$ 2.94  & 90.00 $\pm$ 2.12  & 85.37 $\pm$ 1.90  & 0.03 \\    
        ACS Scratch	& 85.10 $\pm$ 1.79  & 92.82 $\pm$ 0.40  & 70.98 $\pm$ 1.78  & 59.84 $\pm$ 4.12  & 87.81 $\pm$ 3.48  & 84.39 $\pm$ 3.12  & 0.00 \\        
        ACS			& 83.93 $\pm$ 2.07  & 93.54 $\pm$ 0.37  & 70.30 $\pm$ 1.91  & 57.31 $\pm$ 3.90  & 91.56 $\pm$ 4.03  & 81.95 $\pm$ 3.61  & 0.00 \\
        I3D			& 84.77 $\pm$ 1.33  & 93.23 $\pm$ 0.55  & 70.25 $\pm$ 1.24  & 59.16 $\pm$ 3.04  & 86.87 $\pm$ 3.51  & 84.23 $\pm$ 2.47  & 0.00 \\
        Kinetics	& 84.58 $\pm$ 1.26  & 94.04 $\pm$ 0.32  & 70.43 $\pm$ 1.59  & 58.44 $\pm$ 2.52  & 88.75 $\pm$ 1.17  & 83.50 $\pm$ 1.79  & 0.00 \\
        \midrule
        Ours		& 85.49 $\pm$ 0.98  & $\bm{94.84 \pm 0.22}$  & 72.15 $\pm$ 0.79  & 59.92 $\pm$ 2.37  & 90.94 $\pm$ 3.34  & 84.06 $\pm$ 2.05  & $-$ \\
        \bottomrule
    \end{tabular}
    }
\end{table*}


\subsection{Lung Nodule Classification and Segmentation}

\subsubsection{Dataset and Implementation Details}

\noindent{\textbf{Dataset:}}
The Lung Image Database Consortium image collection (LIDC-IDRI) \cite{armato2011lung} is a large scale lung nodule dataset that consists of 2,669 nodules from 1,018 CT scans. Each nodule is annotated by up to 4 radiologists, which has pixel-level mask annotations for lung nodule segmentation and 5-level malignancy scores for malignancy classification. 

Following the prior works \cite{xie2018knowledge,yang2019reinventing}, only nodules with diameters $\geq$ 3mm are adopted in our study, since smaller nodules are not considered to be clinically relevant by current screening protocols. For both tasks, we apply the same data split as used in ACS~\cite{yang2019reinventing} for model training and evaluation. Specifically, for lung nodule segmentation, a total of 2,142 samples are used for training and 526 samples for testing. The Dice coefficient score is reported for model comparison. 

With regards to lung nodule classification, nodules with uncertain annotations (level 3) are further ignored to reduce ambiguity for malignancy evaluation. A binary benign-malignant classification task is conducted by grouping level 1,2 to the benign and level 4,5 to the malignant class. Finally, as done in ACS, we enrolled 1,633 nodules for the lung nodule classification, including 1,156 benign and 556 malignant nodules. We use a series of metrics to evaluate the two-class classification problem, including AUC, accuracy, sensitivity, specificity, precision and F-score for evaluation, among which, the AUC is the main compared metric. 

\noindent{\textbf{Implementation Details:}}
ResNet3D-18 is employed as the backbone for all the compared methods in this experiment, except for Models Genesis, which is based on a 3D U-Net architecture. With regards to nodule segmentation, as done in ACS, to keep a higher resolution for the final output feature maps, we change the stride of the first convolution layer and the third res-block in backbone to 1, and remove the first max-pooling layer. The change in the stride of the convolutional layer would not affect the load of the pre-trained weights. Then, a light-weight FCN-like~\cite{long2015fully} module which consists of two convolution and up-sample layers is employed as the decoder for segmentation map prediction.  
We use ResNet-18 as the backbone for most experiments except Models Genesis, whose official implementation is based on a U-Net-like structure. 

For nodule classification, we use the Adam~\cite{kingma2014adam} optimizer with an initial learning rate ($lr$) of 5$e$-4 to train all compared models for 100 epochs. The learning rate is decreased by a factor of 10 after 30, 60 and 90 epochs respectively.
The polynomial learning rate policy with power of 0.9, initial $lr$ of 1$e$-3 and min $lr$ of 1$e$-6 are used for all nodule segmentation models except Models Genesis with U-Net structure, for which $lr$ is set to 1$e$-4 to achieve its best performance. A dice loss and a cross-entropy loss with a loss weight of $(0.3, 1.0)$ are used for network supervision. Random center-cropping is implemented to generate $48\times48\times48$ image patches as input. We use random center-cropping/flipping/axis-rotation as data augmentation in both classification and segmentation experiments. 

We repeat all experiments by 5 times and report the mean and standard deviation for all evaluation metrics. In each run, we fix the random seed for all compared methods to reduce the difference incurred by random data augmentation.

\subsubsection{Performance Evaluation}


The experimental results of different methods on lung nodule classification task are shown in Table \ref{LIDC-CLS}. 
As done in ~\cite{zhou2020models} and ~\cite{yang2019reinventing}, we also employ AUC as the main compared evaluation metric. It reflects the overall performance of a binary classifier and is not affected by class imbalance problems. On the contrary, accuracy, sensitivity and precision rely on the selection of the threshold and may vary greatly given different thresholds. 

It can be observed that our proposed pre-trained ResNet3D-18 model outperforms competing methods significantly.
Our pre-trained model surpasses the model trained from scratch by $2.09\%$, demonstrating the effectiveness of the proposed pre-training approach. What's more, we achieve better performance than the state-of-the-art self-supervised pre-training approach, Models Genesis, which is pre-trained on a subset of the LIDC dataset. Meanwhile, other natural image pre-trained methods (ACS, I3D, Kinetics) also obtain considerable performance advantage compared with training from scratch, revealing the potential of transferring knowledge learned from large-scale natural image datasets to different domains and tasks. 

Independent two-sample \textit{t}-test is performed between our proposed method and other compared methods with the AUC metric. All the compared pairs are statistically significantly different at $p$=0.05 level, some of them are even significantly different at $p$=0.01 and $p$=0.001 level. These clearly prove that our proposed supervised pre-training framework is able to acquire powerful 3D representations for transfer learning on medical image analysis. 

Table \ref{LIDC-SEG} shows the mean and standard deviation of the dice score of different methods on the LIDC segmentation task.
Our pre-trained model consistently achieves a higher dice score compared with other methods. With the pre-trained weights learned from ImageNet, our nodule segmentation model obtains a dice score improvement of $0.83\%$ compared with training from scratch.


\subsection{Liver Segmentation}

\begin{figure*}
    \centering
	\includegraphics[width=7in]{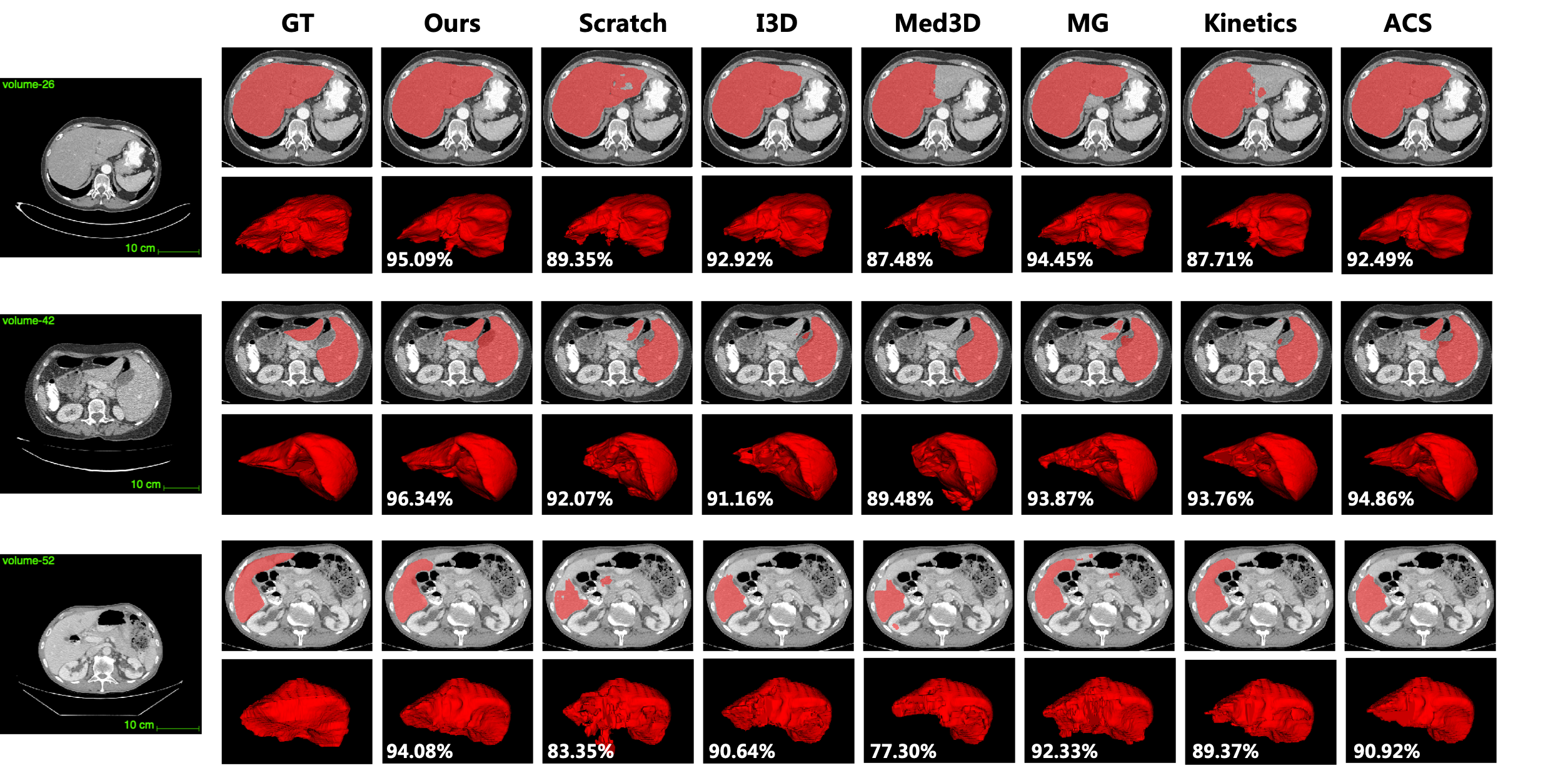}
	\caption{Visual comparison of our proposed methods with competing pre-training methods and training from scratch on the LITS dataset.}
	\label{livervis}
\end{figure*}

\subsubsection{Dataset and Implementation Details}
\noindent{\textbf{Dataset:}}
In this section, we further validate our proposed method on a challenging 3D medical image segmentation dataset. The Liver Tumor Segmentation Benchmark (LiTS)~\cite{bilic2019liver} is composed of 201 contrast-enhanced abdominal CT scans collected from several clinical centers, including 131 volumes for training and volumes for online evaluation. In this experiment, we only use 131 scans whose annotations are publicly available for developing 3D segmentation networks. Segmentation masks of both liver and liver tumors are provided on this dataset. For our experiments, the segmentation target is liver only. The pixel spacing of CT scans ranges from 0.56\textasciitilde1.0 $mm$ in the xy-plane and the slice spacing ranges from 0.45\textasciitilde6.0 $mm$ in z-axis.

\begin{table}
    \centering
    \setlength{\belowcaptionskip}{5pt}
    \caption{Dice Coefficient per Case (DPC) for liver segmentation on LiTS. Results are given as $mean \pm standard \  deviation$ format.}
    \label{LITS-SEG}
    \footnotesize 
    \setlength{\tabcolsep}{11pt}
    \renewcommand{\arraystretch}{1.0}
    \begin{tabular}{lccc}
    \toprule
        Methods & Backbone & DPC & P-value \\ \midrule
        Scratch & ResNet3D-18 & $92.59\pm0.41$ & 0.00 \\         
        Med3D & ResNet3D-18 & $93.71\pm0.54$ &  0.00  \\ 
        Models Genesis & U-Net 3D & $95.68\pm0.14$ & 0.02 \\
        ACS & ResNet3D-18 & $95.64\pm0.37$  & 0.08\\ 
        I3D & ResNet3D-18 & $94.75\pm0.57$  & 0.00\\ 
        Kinetics & ResNet3D-18 & $95.59\pm0.33$ & 0.04\\ \midrule
        Ours (Fix-Res1) & ResNet3D-18 & $95.82\pm0.22$ & 0.15 \\
        Ours & ResNet3D-18 & $\textbf{96.07}\pm\textbf{0.23}$ & -\\ \bottomrule

    \end{tabular}
\end{table}

To remove irrelevant information, we clip the Hounsfield Unit (HU) value to [-200, 250] and then normalize it to [0, 255]. Slice spacing is normalized to 1 $mm$ to mitigate its impact on performance. Following the evaluation protocols of the LiTS challenge, we evaluated the liver segmentation performance with an average of Dice per volume score (Dice per case). 

\noindent{\textbf{Implementation Details:}}
We build a 3D U-Net with ResNet3D-18 as the backbone for liver segmentation. The stride of the first convolution layer is changed to 1 to maintain full image resolution after the stem layers. Models are trained with combined supervision from the dice loss and a cross-entropy loss with a loss weight of $(0.5, 1.0)$. At the training stage, 3D patches with the size of $32\times256\times256$ are cropped out from the normalized CT scans to serve as the network input. For network inference, we densely crop $32\times256\times256$ patches with a sliding stride of $12\times128\times128$ to get the prediction for the whole CT. Data augmentation in the training stage includes random cropping, flipping and re-scaling, and no test-time augmentation is implemented. Adam method with an initial $lr$ of 1$e$-4 is used for optimization. We use polynomial to schedule the learning rate with a power of 0.9 and min $lr$ of 1$e$-6. We randomly split the dataset into training (105 patients), validation (13 patients) and test (13 patients) subsets. We train all the models for 500 epochs. Best performed models on the validation set are used to report the final performance on the test set. Post-processing is implemented by only keeping the largest connected components in the 3D volume. 

In this experiment, we also repeat all experiments by 5 times with fixed random seeds and report mean and standard deviation for all evaluation metrics. It takes approximately three hours to run a ResNet3D U-Net experiment on 8 Titan-XP GPUs.

\subsubsection{Performance Evaluation and Ablation Study}
As shown in Table~\ref{LITS-SEG}, all the pre-training methods significantly improve the dice score compared to training from scratch. Our pre-trained model achieves the best performance of 96.07\%, consistently surpassing all the compared methods (3.48\% higher than training from scratch). We also performed an independent two-sample \textit{t}-test between our proposed method and other compared methods. The superiority of our SVD-Net is statistically significant ($p<0.05$) for almost all competing methods, except for ACS (marginally significant, i.e. $p<0.1$). Visual comparison of the segmentation results on three CT volumes from the test set is provided in Figure~\ref{livervis}.

Among the competing methods, Models Genesis, which adopts the in-domain medical data for feature learning, achieves the best performance. 
It should be noted that the UNet-3D network used by Models Genesis is slightly larger than the ResNet3D U-Net employed by all other methods.
To further study the impact of domain difference on the performance of transfer learning, we conduct an ablation study on our proposed SVD-Net. 
Specifically, by keeping the pre-trained weights on the stem layers and the first ResBlock fixed during fine-tuning on the target task (i.e. Ours (Fix-Res1)), we achieve a dice score of 95.82\%, which still outperforms all the competing methods. And the difference between Ours and Ours-Fix-Res1 is not significant (p-value $>$ 0.1). 
This indicates that shallow features learned from our pseudo-3D data can be well generalized to medical data without any further fine-tuning, validating our motivation for learning representations from cross-domain data. 

\subsection{Universal Lesion Detection}

\subsubsection{Dataset and Implementation Details}

\noindent{\textbf{Dataset:}}
The NIH DeepLesion is a large-scale dataset for universal lesion detection, which contains 32,735 lesions on 32,120 axial CT slices captured from 4,427 patients. RECIST diameter coordinates and bounding boxes are provided on the key slices, with adjacent slices (above and below 30$mm$) also provided as contextual information. Only bounding box annotations are used in this experiment. The official split of DeepLesion with training (70\%), validation (15\%), and test (15\%) sets are used in our experiments. We evaluate all the compared methods on the test set by reporting sensitivities at different false positives (FPs) per image and their average (mFROC). As for pre-processing, the Hounsfield units (HU) are clipped into the range of $[-1024,1050]$. We also implement interpolation in the z-axis to normalize the intervals of all CT slices to 2.5mm.

\noindent{\textbf{Implementation Details:}}
All the compared methods take 9 consecutive slices as input, which can be represented as a gray-scale 3D tensor of $1\times9\times512\times512$. Since only 2D annotations (2D bounding boxes on the key slice) are provided in the DeepLesion dataset, we use a similar architecture as MP3D~\cite{zhang2020revisiting} for lesion detection. MP3D neglects all the down-sampling operations for the z-axis to keep the depth of all backbone outputs as 9. In our implementation, to further improve model efficiency, we conduct down-sampling for the z-axis like the original ResNet architecture on all the res-blocks, which will result in 3D feature maps of size $1 \times (9, 5, 3, 1, 1) \times H \times W$ for each FPN level. ResNet3D-18 is adopted as the backbone network for comparison with competing pre-trained methods. 


Anchor scales of FPN are set to $(16, 32, 64, 128, 256)$ to improve detection performance for small lesions. We apply multi-scale training with scales randomly sampled from $(384, 448, 512, 576, 640)$. Apart from re-scaling, we also implement horizontal and vertical flipping for input tensors during training, and no other data augmentation strategies are applied. No test-time augmentation (TTA) is applied for model inference. All models are trained for 24 epochs at the base learning rate of 0.02 with SGD, and reduce it by a factor of 10 after the 16-th and 22-th epoch (corresponding to the 2x learning schedule\cite{he2019rethinking} on MS-COCO). 


\subsubsection{Performance Evaluation}
Table \ref{SOTA} depicts the performance of previous state-of-the-arts and compared pre-trained models. With our proposed pre-training technique and a P3D-63~\cite{hara2017learning} backbone, we achieve a new state-of-the-art mFROC of 88.55\% on the DeepLesion Dataset, surpassing previous state-of-the-art by a large margin. Note that unlike previous state-of-the-arts AlignShift, which also take advantage of RECIST supervision and tags from medical reports and demographic information, our detection models only use bounding box annotation for lesion detection. The P3D-63 backbone has a similar model capacity with ResNet3D-50 and DenseNet3D-121, but it is more computational- and memory-efficient with its specially designed model architecture. 

\begin{table*}[!t]
    \setlength{\belowcaptionskip}{5pt}
    \caption{Sensitivities (\%) at various FPs per image on the test set of NIH DeepLesion. Ours (ImageNet) and Ours (MS-COCO) indicates pre-training on ImageNet and MS-COCO respectively.}
    \label{SOTA}
    \centering
    \footnotesize 
    \setlength{\tabcolsep}{11pt}
    \renewcommand{\arraystretch}{1.2}
        \centering
        \label{tab:sample_1}
        \begin{tabular}{lcccccccccc}
            \toprule
            \textbf{Methods}  & \textbf{Backbone} & \textbf{Slices}  & $\mathbf{0.5}$ & $\mathbf{1}$ & $\mathbf{2}$ & $\mathbf{4}$ & $\mathbf{mFROC[0.5,1,2,4]}$ & \\             \midrule
            3DCE~\cite{yan20183d}  &  ResNet-50 & 27 slices &  52.86 & 64.80 & 74.84 & 84.38 & 69.22  \\
            MVP-Net~\cite{li2019mvp}  & ResNet-50 & 9 slices & 73.83 & 81.82 & 87.60 & 91.30 & 83.64 \\    
            ACS~\cite{yang2019reinventing}  & DenseNet3D-121 & 7 slices & 78.38 & 85.39 & 90.07 & 93.19 & 86.76 \\
            AlignShift~\cite{yang2020alignshift}  & DenseNet3D-121 & 7 slices & 79.40 & \textbf{85.50} & \textbf{90.09} & \textbf{93.26} & \textbf{87.06} \\
            MP3D~\cite{zhang2020revisiting}  & MP3D-63 & 9 slices &  \textbf{79.60} & 85.29 & 89.61  & 92.45  & 86.74  \\   \midrule         
            Scratch & ResNet3D-18 & 9 slices & 66.51 & 74.20 &  80.33 &  85.22 &  76.57 \\ 
            Med3D & ResNet3D-18 & 9 slices & 58.66 & 68.36 & 76.11 & 82.46  & 71.40  \\
            ACS & ResNet3D-18 & 9 slices & 71.16 & 78.95 & 84.98 & 89.20  & 81.07 \\
            I3D & ResNet3D-18 & 9 slices & 57.93 & 67.51 & 75.54 & 81.69 & 70.67 \\
            Kinetics & ResNet3D-18 & 9 slices & 72.94 & 80.92 & 86.00 & 89.91 & 82.44 \\
            Ours (ImageNet) & ResNet3D-18 & 9 slices & \textbf{74.18} & \textbf{81.55} & \textbf{86.77} & \textbf{90.68} & \textbf{83.30} \\\midrule
            Ours (MS-COCO) & ResNet3D-18 & 9 slices & 76.07 & 82.16 & 86.67 & 90.12 & 83.76 \\ 
            Ours (MS-COCO) & ResNet3DV1c-18 & 9 slices & 79.28 & 84.8 & 89.04 & 91.90 & 86.26 \\
            Ours (MS-COCO)  & P3D-63 & 9 slices &  \textbf{82.22} & \textbf{87.42} & \textbf{90.91} & \textbf{93.65} & \textbf{88.55}   \\ \bottomrule
        \end{tabular}
\end{table*}

Compared to the MP3D, our implementation features the down-sampling operation for z-axis on the 3D backbone. Such an implementation can significantly reduce the computational cost for feature extraction, and is a natural fit for the FPN architecture. With regards to FPN levels that are adopted to detect small-sized bounding boxes, the feature maps has larger resolutions on the z-axis (9,5,3,1,1 for each level), and vice versa. Therefore, the down-sampling operation will not impose a negative impact on the localization performance. 

\begin{table}[!t]
    \setlength{\belowcaptionskip}{5pt}
    \caption{Ablation studies of pre-train architecture on NIH DeepLesion dataset. w/o mod indicates models trained without proposed modifications, i.e. the models are trained with z-pooling on the depth axis. All compared models are pre-trained on the MS-COCO dataset.}
    \label{ablation}
    \centering
    \footnotesize 
    \setlength{\tabcolsep}{4pt}
    \renewcommand{\arraystretch}{1.2}
        \centering
        \begin{tabular}{lcccccc}
            \toprule
            \textbf{Methods} & \textbf{Backbones} &  $\mathbf{0.5}$ & $\mathbf{1}$ & $\mathbf{2}$ & $\mathbf{4}$ & \textbf{mFROC} \\
            \midrule
            Ours & ResNet3DV1c-18 & 79.28 & 84.8 & 89.04 & 91.9 & \textbf{86.26}  \\
            w/o mod & ResNet3DV1c-18 &  76.07 & 82.16 & 86.67 & 90.12 & 83.76  \\ \bottomrule
        \end{tabular}
\end{table}

From the second part of Table \ref{SOTA}, we can observe that our proposed pre-trained weights consistently outperforms all the competing pre-training method and training from scratch. Till now, our SVD-Net has been able to achieve better performance on all three types of target medical image analysis tasks, e.g. classification, segmentation and detection.

\subsubsection{Ablation Study on Pre-training Task and Architecture}
In this subsection, we conduct three ablation studies to explore the impact of pretext tasks and model architectures. In Table \ref{SOTA}, Ours (MS-COCO) slightly outperforms Ours (ImageNet) on the mFROC for our target detection task. This suggests that task similarity between the pretext task and the target task should be taken into consideration when selecting a proper pre-trained model.

We also look into the details in network design for performance ablation. Normal ResNet3D-18 adopts a $7\times7\times7$ kernel for the stem convs, however, when pre-training with our proposed pseudo-3D input, its depth is only three. Large portions of the convolutional kernels will not be updated during pre-training. To mitigate this problem, we adopt a network architecture proposed in~\cite{he2019bag}, i.e. ResNet3DV1c-18 for pre-training. The ResNet3DV1c-18 model replaces the $7\times7\times7$ kernel with three $3\times3\times3$ kernels, so that all the parameters will be able to update during pre-training. Experimental results in Table \ref{SOTA} demonstrate that such a change greatly improves model performance for the lesion detection task. This suggests that proper training of the 3D kernels is essential for advanced performance on the target task.

In Table~\ref{ablation}, we further conduct an experiment to validate the importance of 3D kernel learning in our proposed pre-training framework. Specifically, we pre-train a ResNet3DV1c-18 backbone on the MS-COCO dataset without our proposed modifications, i.e. the models are trained with vanilla down-sampling operations on the depth dimension. With such a model, the depth dimension will be down-sampled to 1 after the first two down-sampling layers, and thus 3D kernels in the succeeding layers will be degenerated to 2D ones, since only the center $3\times3$ part could be optimized. When applying those two pre-trained models on the lesion detection task, a large performance gap (2.50\% on mFROC) is observed. This evidently proves that our proposed modification to the pre-trained network plays a vital role for effective learning of 3D representations.

\subsubsection{Performance with Limited Training Data}

To evaluate the robustness of the proposed method in data-limited scenarios, we train a series of models with various amount of annotated training data (20\%, 40\%, 60\%, 80\% and 100\%) under the same experimental setting. 

\begin{figure*}
    \centering
 	\includegraphics[width=\textwidth]{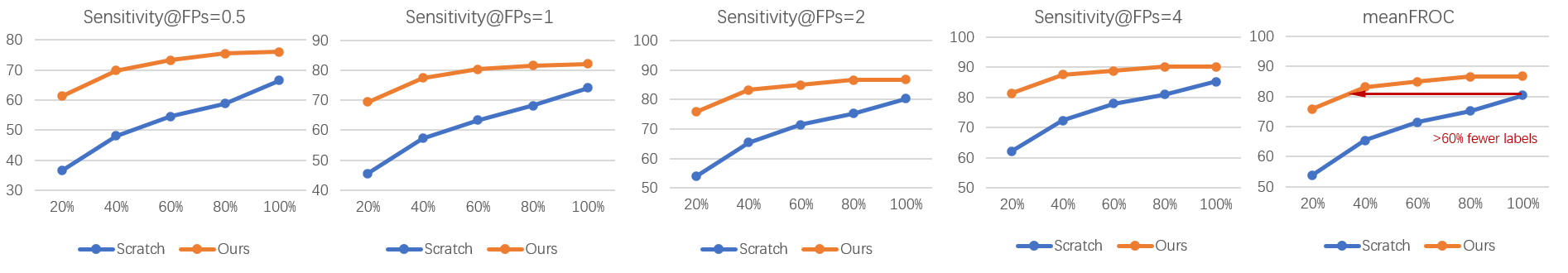}
	\caption{Detection performance on the NIH DeepLesion dataset at different annotated data scales.}
	\label{data_scale_cls}
\end{figure*}

In Figure~\ref{data_scale_cls}, it can be clearly seen that compared with training from scratch, our proposed pre-trained models consistently obtain better performance. And the performance gap is larger when fewer training samples are available (the gap on mFROC is 22.44\% when training on 20\% samples, while when training on 80\% samples, the gap is 12.64\%). This indicates that when dealing with small-scale datasets, the pre-trained weights have a larger influence on the final performance. From the last diagram in Figure~\ref{data_scale_cls}, we can observe that using only about 38\% of the annotated data, our pre-trained models achieve equal performance to that of training from scratch with all the annotated data. Therefore, the cost of annotation could be reduced by over 60\% when initializing 3D models with our proposed pre-trained weights compared with training from scratch. This indicates that with our proposed pre-training framework, we will be able to efficiently verify some initial ideas with fewer annotation labors. This would promote the development of 3D-based deep learning methods for 3D medical image analysis problems.

\section{Conclusions}
\label{con1}
In this paper, we propose a novel variable dimension transform based fully-supervised learning framework to pre-train 3D neural networks for 3D medical imaging tasks. The proposed SVD-Net takes advantage of the large-scale semantic annotations of 2D natural image dataset to enforce learning of discriminative and invariant 3D feature representations with the proposed variable dimension transform scheme. Compared to self-supervised methods, the learned representations enforced by semantic annotation are more general and discriminative. Although there exists a certain degree of domain shift between the reformulated natural image and 3D medical image, comprehensive empirical experimental results demonstrate that our proposed method outperforms previous state-of-the-art 3D pre-training methods on four benchmark datasets for 3D medical image analysis, validating its strong ability in generalizing to major tasks like classification, segmentation and detection. 
Moreover, comprehensive experiments are designed and analyzed to look into the factors that contribute to the success of our proposed methods. 
To benefit the research community, we release our codes and pre-trained weights to the public.

\bibliographystyle{IEEEtran}

\end{document}